\documentclass[a4paper,11pt,twoside]{article}
\def\supit#1{\raisebox{0.8ex}{\small\it #1}\hspace{0.05em}}  
\def\skiplinehalf{\medskip\\}
\newcommand{\ket}[1]{|#1\rangle}
\title{Evolution of a qubit under the influence of a succession of unsharp
measurements}
\author{J\"urgen Audretsch\supit{a}
 Lajos Di\'osi\supit{b} and Thomas Konrad\supit{a}
\skiplinehalf 
\supit{a}
Fachbereich Physik,
Universit\"at Konstanz, Fach M 674,\\
D-78457 Konstanz, Germany
\\
\supit{b}
Research Institute for Particle and Nuclear Physics, \\
H-1525 Budapest 114, P.O.Box 49, Hungary
\skiplinehalf 
 }
\date{18.01.2002}

\begin{document}
\maketitle
\begin{abstract} We investigate the evolution of a single qubit subject to a
continuous unitary dynamics and an additional interrupting influence
which occurs periodically.
One may imagine a dynamically evolving closed quantum
system which becomes open at certain times. The interrupting 
influence is represented by an operation, which is assumed to 
equivalently describe  a non-selective unsharp measurement. It may be
decomposed into a positive operator, which in case of a measurement
represents the pure measurement part, followed by an unitary
back-action operator. Equations of motion for the state evolution are
derived in the form of difference equations. It is shown that the
\lq\lq free" Hamiltonian is completed by an averaged Hamiltonian, which goes
back to the back-action. The positive operator specifies a decoherence
rate and results in a decoherence term. The continuum limit to a
master equation is performed. The selective evolution is discussed and
correcting higher order terms are worked out in an Appendix.
\end{abstract} 

\section{Introduction}
Experimental and theoretical studies of the dynamics of single
two-level systems have become very important in the context of
quantum computation and quantum information. In this article we
investigate the evolution of a single qubit subject to a
continuously acting unitary dynamics (undisturbed or \lq\lq free"
dynamics) with operator $U$ as well as affected by an
interrupting additional influence, which is non-unitary and acts
periodically at times $t_n= t_0+n\tau$, $n= 1\,,\,2\,...$. The
duration $\delta \tau$ of this influence is assumed to be much
shorter than $\tau$, so that it can be neglected. One may imagine
a dynamically evolving closed quantum system which becomes open at
times $t_n$. The corresponding single influence is represented by
an {\sl operation} $\mathcal{E}$ which transforms the state
 of the qubit given by its reduced density  operator $\rho$
according to
\begin{equation}
\label{operationE}
\rho\rightarrow \mathcal{E}(\rho)=\sum_{k=\pm}\,M_k\rho M_k^+
\end{equation}
with {\sl operation elements} $M_k$ , which are sometimes also
called Kraus operators. $\mathcal{E}$ is assumed to be trace
preserving. The representation (\ref{operationE}) of the single
operation is called the {\sl operator-sum representation} or
Kraus representation.

Such a periodically occurring, nearly instantaneous change can be
caused by a recurring interaction with a second system provided
that this system does not \lq\lq remember'' the influence it may
have experienced from the qubit at former times (Markov process).
Typically the second system could be an environment or it could
consist of a number of systems of the same kind which interact
only once with the qubit, as it is the case in a sequence of
scattering processes.

According to eqn.\ (\ref{operationE}) we  are not dealing with the
most general form of such an influence on a qubit, which would
correspond to an operation with four operation elements, but we
restrict to interactions which may be represented by only two
elements $M_+$ and $M_-$.  Pairs of operation elements can
describe such important operations as for example amplitude
damping and phase damping, bit flips and phase flips as well as
projection measurements and unsharp measurements. The concept of
an unsharp measurement will be explained below. We will restrict
to an operation which is equivalent, as far as the map
$\mathcal{E}$ is concerned, to a non-selective unsharp
measurement. Note that this may still comprise many physical
processes which at first glance do not look like an unsharp
measurement. An example is the experiment of Brune, Haroche et
al. \cite{Brune.et.al90} to measure the number of photons in a
cavity, which is discussed under the aspect of an unsharp
measurement in \cite{AudretschKonradScherer02}.

According to the polar decomposition theorem each operation
element $M_\pm$ may be written as a product of a unitary operator
and a positive operator
\begin{equation}
M_\pm = U_\pm|M_\pm|\,.
\end{equation}
We are dealing with a class of generalized measurements with two
outcomes $+$ and $-$. The unitary part $U_\pm$ extracts no
information from the qubit. It is in a different context often
called the feedback part of the quantum operation
\cite{DohertyJacobsJungman00, Wiseman95}. The set
$\{|M_-|^2,\,|M_+|^2\}$ represents a positive-operator valued
measure (POVM). The demand of unsharpness of the measurement is
introduced below in specifying these $|M_\pm|$ further. A sequence
of such generalized measurements are of practical importance
because they can be employed to explore the original dynamics of
the system \cite{AudretschKonradScherer02,
AudretschKonradScherer01} or to control its dynamics by means of
a specific feedback \cite{DohertyJacobsJungman00,
Doherty.et.al00}.

In the non-selective case of eqn.\ (\ref{operationE}), when the
measurement results $\pm$ are not read off, our total physical
setup defined by $U, \tau, U_{\pm}$ and $M_\pm$ may also be regarded
as a particular noisy channel. Below the unitary operators $U_\pm$
will not be neglected because they form an important part of the
sequential operations in realistic situations. In general it would
need a non-trivial feedback procedure to eliminate their
influence, compare \cite{AudretschKonradScherer02} for an example.

It is our goal to derive equations of motions for the state of the
system subject to the periodical influence. We will do this in
form of difference equations and -- on a coarse grained time scale
-- in form of master equations.  Difference equations take into
account the discrete nature of the influence due to the finite
time $\tau$ between its occurrences and grant therefore a more
exact description. It may reveal specific physical influences
which cannot be seen in the continuum limit furthermore.

We also look at the limit $\tau\rightarrow 0 $ of continuous
measurements and compute the corresponding master equation
(non-selective description). The master equation can be
understood as a means to compute approximately the dynamics of a
system which is subject to a sequence of operations of the
type (\ref{operationE}). Since there are a lot more mathematical
methods to solve differential equations than difference
equations, it is often useful to approximate sequential
measurements by continuous measurements.

Master equations for special cases of measurements with
non-minimal disturbance of the state , i.e. with a non-vanishing
unitary part of the operation elements have been considered in
the literature. A master equation for general feedback was
derived by Wiseman \cite{Wiseman94}. In the Markovian limit, if
the time delay between feedback and measurement vanishes and the
feedback depends only on the outcome of the last measurement
(instantaneous feedback), the action of the feedback can be
represented as unitary part of the operation of the measurement.
However \cite{Wiseman94} does not comprise our results since it
deals with a special kind of continuous measurements. They have
poissonian statistics and  allow finite state changes during
infinitesimal time intervals. More precisely this means  that
only very seldom a certain measurement result occurs which is then
connected to a finite state change during a infinitesimal time
while for other  measurement results the state changes only
infinitesimally. We are excluding Poissonian statistics and
require the state change  in the continuum limit to be
infinitesimal during infinitesimal times. Thus Wiseman's and our
studies do not overlap.

In a later paper Wiseman  \cite{Wiseman95} employed
the operation formalism to analyze a homodyne measurement in quantum
optics to apply instantaneous feedback
in order to minimize disturbance, i.e., compensate the
unitary part of the operation. Korotkov  investigated a
measurement with non-minimal disturbance in the context of continuous
measurement of a qubit by means of a single electron
transistor \cite{Korotkov00a,Korotkov00b}. He
noted that this non-minimal disturbance acts  in the master equation
like a change of the distance between the two energy levels of the
qubit. We find this effect as a special case of our studies
($[U_\pm,H]=0$). Korotkov also derives a modification of the
stochastic master equation (selective regime)
due to the non-minimal disturbance \cite{Korotkov00a}, which is
averaged out in the non-selective case.

Many examples of the application of instantaneous feedback in
continuous measurements by means of
external changes of the Hamiltonian of the system can be found in the
literature of quantum control (e.g. \cite{DohertyJacobsJungman00, Doherty.et.al00}). They correspond to
special choices of the unitary part of the measurement operation and
obey the equations of motions derived here, provided the measurements
do not inflict finite changes during infinitesimal times.

In the context of quantum dissipation the influence of a heath
bath on a infinite dimensional quantum system has been
investigated by Caldeira and Leggett \cite{CalderaLeggett83}. The
coupling was such that in terms of the  operation formalism the
corresponding operation of the system had a unitary part
additional to the one stemming form its free original evolution.
They derived  a master equation for high temperatures which was
later modified to also describe medium temperatures
\cite{Diosi93}. Although we are looking at a qubit we find in the
difference equation among others similar terms like there, but
only one of them survives in the continuum limit.

In order to discuss the problem in an illustrative way but without
restriction of generality, we treat the periodical influence 
in terms of a sequential measurement. We proceed as follows. We
first consider the single measurements of the sequence. Then we
bundle the whole sequence in subsequences of N measurements
(\lq\lq N-series``). The resulting operation has Gaussian shape.
Afterwards we integrate over all outcomes  to obtain the average
state change due to a N-series (non-selective regime). We then
discuss how to find the right continuum limit which conserves the
physical characteristics of the N-series and derive the master
equation. Finally we deal with the selective regime of
measurement and write down the stochastic master equation. The
Appendices serve to derive the difference equation for the
non-selective regime up to second order in the occurring small
parameters.

\section{The single quantum operation}\label{sectwo}
An example for a physical realisation of the operation $\mathcal{E}$ from
(\ref{operationE}) is given by  a qubit which interacts at times $t_n$
unitarily for a short duration $\delta \tau$ with an environment and
thus becomes an open system. The resulting change of its reduced 
density operator may
formally be expressed with the help of operation elements $M_\pm$
as in eqn. \ (\ref{operationE}). Just the same change of the reduced
density operator results if i) a projection measurement on the
environment is performed with outcome $+$ or $-$ transferring the
qubit to the states
\begin{equation}
\label{state transformation} \rho\,\longrightarrow\,\rho_\pm =
M_{\pm}\rho M_{\pm}\,.
\end{equation}
respectively and ii) the outcome is not read off (non-selective
case). In the generic case, eqn. \ (\ref{state transformation})
describes thereby a generalized measurement of the qubit. For
simplicity reasons the terminology we are going to use will refer
to measurements, but the results apply equally to any operation
with operator-sum representation (\ref{operationE}) if the same
specifications of $M_\pm$ are made. This is independent of how
the operation is experimentally realized.

Because of the polar decomposition theorem, the operation
elements $M_\pm$ may be written as products of a unitary operator
and a positive operator
\begin{equation}
\label{operations} M_{\pm}=U_{\pm}|M_{\pm}|\,.
\end{equation}
We introduce the POVM {\sl effects}
\begin{equation}
\label{effects} E_{\pm}= |M_{\pm}|^2\,,
\end{equation}
which obey the completeness relation
\begin{equation}
\label{povm} E_{+}+E_{-}= 1\hspace{-1.57mm}1\,.
\end{equation}
The probability of the outcome $+$ or $-$ is given by
\begin{equation}
\label{probabilities_1}
p_{\pm}=\langle\, E_{\pm}\,\rangle_{\rho}
\end{equation}
with $p_+ + p_-=1$.

Eqn.\ (\ref{operations}) represents a decomposition of the
operation into a {\sl pure  measurement part} described by
$|M_\pm|$, followed by a unitary {\sl back-action} given by
$U_\pm$ depending on the result $+$ or $-$. These denominations
are justified for the following reasons: All the information which
can be read off from the meter 
is related to $|M_\pm|$ which therefore represents the
unavoidable minimal disturbance. The unitary operators leave the
von Neumann entropy unchanged and therefore do not allow to export
information to an observer. Because they depend on the result $+$
or $-$, they may be interpreted as a specific back-action of the
measuring apparatus inducing an additional
Hamiltonian evolution of the qubit. We formally introduce the
corresponding Hamiltonians $H_\pm$ according to
\begin{equation}
\label{Hplusminus} U_\pm =:
\exp\left(-\frac{i}{\hbar}H_\pm\tau\right)\,.
\end{equation}
This unitary back-action represents an important part of the
quantum operation and appears naturally in the generic situation.

According to eqns.\ (\ref{effects}) and (\ref{povm}) $|M_+|^2$ and
$|M_-|^2$ commute. Therefore we can find orthonormal basisvectors
$|1\rangle$ and $|2\rangle$ of the qubit Hilbertspace with respect to
which $|M_\pm|^2$ are diagonal. We introduce the eigenvalues $p_1$ and
$p_2$ of $|M_+|^2$, which are positive and because of
eqn. (\ref{povm})
obey $0\le p_{1,2}\le 1$. Without restriction of generality we choose
$p_2\ge p_1$. Reading off the eigenvalues of $|M_-|^2$
from (\ref{povm}) and
taking the square root we find
\begin{eqnarray}\label{minimaloperations}
|M_+|&:=&\sqrt{p_1}\,|1\rangle \langle 1|+\sqrt{p_2}\,| 2 \rangle \langle
2|\;\\
|M_-|&:=&\sqrt{1-p_1}\,|1\rangle \langle 1|+\sqrt{1-p_2}\,| 2 \rangle \langle
2|\nonumber
\end{eqnarray}
The elements $|M_+|$ and $|M_-|$
commute. We will characterize the operation later on by the
parameters
\begin{equation}
p_0:=\frac{1}{2}(p_1+p_2)\;,\;\Delta p:=p_2-p_1
\end{equation}
with $0\le \Delta p \le 1$. Introducing
\begin{equation}
\sigma_z:=|1\rangle\langle1|-|2\rangle\langle2|
\end{equation}
 the effects $E_{\pm}$ of eqn.\ (\ref{effects}) are rewritten in the form
\begin{equation}
\label{effect2} E_{+}= p_0 1\hspace{-1.57mm}1 - \frac{1}{2}\Delta
p \sigma_z,~~~ E_{-}= (1-p_0) 1\hspace{-1.57mm}1 +
\frac{1}{2}\Delta p \sigma_z\,.
\end{equation}

In the limiting case $\Delta p=1$, the pure part of the 
measurement (\ref{minimaloperations}) results in a projection on
$\ket{1}$ or $\ket{2}$ depending on the measurement outcome $+$
or $-$. We call this a {\sl sharp measurement} of an observable
with eigenvectors $\ket{1}$ and $\ket{2}$, for example
$\sigma_z$. Note that also for a sharp measurement the result of
the quantum operation (\ref{operationE}) will in general not be
the state $\ket{1}$ or $\ket{2}$ because of the remaining
influence of the unitary back-action.

In the contrary limit $\Delta p\ll 1$ the $|M_+|$ and $|M_-|$
are nearly proportional to the identity operator. The probability
$p_+$ (or $p_-$) to obtain the measurement result $+$ (or $-$) is
then nearly independent of the initial state of the qubit. There
is almost  no state discrimination. Because of this low sensitivity we
call this an {\sl unsharp measurement}. Note that in this limit
the parameters $p_0$ and $1-p_0$ become, because of eqns.
(\ref{probabilities_1}) and (\ref{effect2}), approximately the mean
probabilities to obtain the measurement results $+$ or $-$
respectively. \lq\lq Unsharpness" does of course not originate from a
measurement apparatus which is \lq\lq broken".

We are now able to further specify  the particular class of
quantum operations (\ref{operationE}) which we are going to
discuss. We will restrict to the case that the  pure measurement
part represents an unsharp measurement: $\Delta p\ll 1$.

We will mainly be interested in the non-selectiv case where an
information about the results $\pm$ or the corresponding states
of the environment is not available. The influence on the qubit
at times $t_n$ may then be written in the operator-sum
representation as in eqn.\ (\ref{operationE}).
\begin{equation}
\rho\rightarrow \mathcal{E}(\rho)=\sum_{k=\pm}\,M_k\rho M_k^+\,,\nonumber
\end{equation}
whereby
\begin{equation}
\label{sum}
\sum_{k=\pm}\,M_k^+M_k=1\hspace{-1.57mm}1\,.
\end{equation}
because of (\ref{povm}). The quantum operations at times $t_n$ are
trace preserving.

\section{N-series and related operation}
The time between  measurements is $\tau$. We assume that the
duration $\delta \tau$ of a measurement is much shorter than
$\tau$. The undisturbed or  \lq\lq free" dynamics of the system between
the measurements is given by the Hamiltonian $H$. We bundle $N$
consecutive measurements to a {\sl N-series} of duration $\Delta
t =N\tau$ as we have done in \cite{AudretschKonradScherer02} (cp. also
\cite{AudretschMensky98}).
This procedure has several advantages. We will obtain effects of a
Gaussian structure. This enables us to work out the operator sum
explicity. A comparison with the results in the literature
regarding continuous measurements becomes more evident. And
finally the discussion of the selective case is simpler.

We  require
\begin{equation}
\label{condN}
N\gg 1\,.
\end{equation}
We relate $N$ to the sharpness $\Delta p \ll 1$ of the
measurement by
\begin{equation}\label{eqN}
N\cdot \Delta p \ll 1
\end{equation}
and demand in addition
\begin{equation} \label{condH} \Delta t
\Vert H \Vert \ll \hbar
\end{equation}
and
\begin{equation}
\label{Hfeedsmall}
\Delta t \Vert H_\pm \Vert \ll \hbar\,.
\end{equation}
This means that the influence of the undisturbed dynamics of the
qubit and the unitary back-action dynamics due to the measurements
are both small over the duration $\Delta t$ of a N-series. With
$\Delta t = N\tau$ we have obtained above restrictions for $N,
\tau, H$ and $H_\pm$.

The density operator resulting at the end of a N-series of
measurements with results $m_1,\,...\,m_N$, each of which can
assume the values \lq\lq +" and \lq\lq $-$", read
\begin{equation}
\label{N_series}
\rho(t+\Delta t) =
M_{m_N}U\,...\,M_{m_1}U\rho(t)U^+
M_{m_1}^+\,...\,U^+ M_{m_N}^+
\end{equation}
with
\begin{equation}
\label{unitary_dynamics}
U:=\exp{\{-\frac{i}{\hbar}H\tau\}}\,.
\end{equation}

The influences of the systems dynamics and the measurement will in
general not commute. The following relation is derived in
Appendix A:
\begin{equation}
\label{C1}
M_{m_N}UM_{m_{N-1}} U\,...\,M_{m_1} U= M_{m_N}M_{m_{N-1}}
\,...\,M_{m_1} U^N(1 +C_1)
\end{equation}
with
\begin{equation}
\label{NormC1} \Vert C_1\Vert \le \mathcal{O}(N\Delta p \Delta t
\Vert H \Vert /\hbar) + \mathcal{O}(\Delta t^2 \Vert H \Vert
\max\{\Vert H_\pm\Vert\}/\hbar^2)\,.
\end{equation}
Also the operations $M_\pm$ will not commute. Based on the
decomposition (\ref{operations}) we show in Appendix A that
\begin{equation}
\label{C2}
U_{m_N}|M_{{m_N}}|\,...\,U_{m_1}|M_{m_1}| U^N= U_+^{N_+}U_-^{N-N_+}
|M_+|^{N_+}|M_-|^{N-N_+}U^N(1 +C_2)\,.
\end{equation}
with
\begin{equation}
\label{NormC2} \Vert C_2\Vert \le \mathcal{O}(N\Delta p \Delta t
\Vert \max\{\Vert H_\pm\Vert\}/\hbar) + \mathcal{O}(\Delta t^2
\Vert H_+\Vert\Vert H_-\Vert/\hbar^2)\,.
\end{equation}
$N_+$ and $N-N_+$ are the total numbers of measurement results
\lq\lq$+$" and \lq\lq$-$" in the N-series respectively. Because of
the assumptions (\ref{eqN}), (\ref{condH}) and (\ref{Hfeedsmall}) we
may neglect $C_1$ and $C_2$. The calculation which takes into
account $C_1$ and $C_2$, is given in the Appendix C.

In our approximation the influence (\ref{N_series}) of the
N-series becomes a function of $N_+$ only, independent of the
${N\choose N_+}$ different orderings of the \lq\lq$+$" and
\lq\lq$-$" results (cp. eqn.\ (\ref{C2})). Therefore the total
N-series of duration $\Delta t$ including the \lq\lq free" dynamics
corresponds to a quantum
operation with operation elements
\begin{equation}
\label{M}  M(N_+,N)=U_+^{N_+}U_-^{N-N_+}|M(N_+,N)| \exp\Big\{ -
\frac{i}{\hbar} H \Delta t \Big\}
\end{equation}
with
\begin{eqnarray}
\label{MNplus}
|M(N_+,N)|&=&\sqrt{{N\choose N_+}}|M_+|^{N_+}|M_-|^{N-N_+}\\
&=&\sqrt{{N\choose N_+}}\,\Big(\,p_1^{\frac{N_+}{2}}
(1-p_1)^{\frac{N-N_+}{2}}\,|1\rangle \langle 1|+
p_2^{\frac{N_+}{2}}
(1-p_2)^{\frac{N-N_+}{2}}\,|2\rangle \langle 2|\,\Big)\,.\nonumber
\end{eqnarray}
The square root in front ensures the completeness relation of the
effects
\begin{equation}
\label{complete}
\sum_{N_+=0}^{N}\, M(N_+,N)^+M(N_+,N)=1\,.
\end{equation}
The unitary back-action part caused by the N-series measurements
can be expressed as
\begin{eqnarray}
\label{UN}
U(N_+,N)&:=&U_+^{N_+}U_-^{N-N_+}\nonumber\\
        &=&\exp{\left\{-\frac{i}{\hbar}\Big(N_+H_++(N-N_+)H_-\Big)\tau\right\}}\,.
\end{eqnarray}

We now make use of the condition that $N$ is large, so that
$|M(N_+, N)|$ of eqn.\ (\ref{MNplus}) may approximately be written
in form of a Gaussian
\begin{equation}
\label{MN} |M(N_+,N)|=\frac{1}{\sqrt[4]{2\pi N E_+ E_-}}
\exp\bigg\{-\frac{(NE_+ -N_+)^2}{4NE_+E_-}\bigg\}\,,
\end{equation}
which contains the operators $E_\pm$ of (\ref{effect2}). Because
we assumed that the measurements are unsharp and therefore
$\Delta p \ll 1$, the \lq\lq spread" of the Gaussian becomes in lowest
order a c-number
\begin{equation}
\label{EplusEmin} E_+E_-=\Big(p_0(1-p_0)-\frac{1}{4}(\Delta
p)^2\Big)1\hspace{-1.57mm}1 +\Delta
p\big(p_0-\frac{1}{2}\big)\sigma_z = p_0(1-p_0)\,,
\end{equation}
where we have ignored terms of order $\Delta p$ and higher on the
right-hand side of (\ref{EplusEmin}). The error thus committed in the
Gaussian in (\ref{MN}) is of order $N\Delta p^3$, which can be seen by
inserting $E_+$ from (\ref{effect2}) and $N_+/N$ from (\ref{defs}) and
expanding Gaussian (\ref{MN}) in powers of $\Delta p$. A more detailed 
calculation can be found in
Appendix C.

We introduce a new variable $s$ to replace the readout $N_+/N$
according to 
\begin{equation}
\label{defs} \frac{N_+}{N}=: p_0-\frac{1}{2}\Delta p s\,.
\end{equation}
Because $N$ is large we may approximately regard $s$ to be
continuous. Its range is limited by:
\begin{equation}
\label{lims} 0\leq p_0 -\frac{1}{2}\Delta p s \leq 1\,.
\end{equation}

%The  continuity of the
%possible outcomes $s$ results from the unsharpness of the measurement.
%(??? Such a measurement can be also realized with a single measurement
%instead of an N-series by coupling our qubit system to
%an infinite dimensional ancilla and carrying out a ideal measurement
%of a continuous observable of the ancilla. ???)

In addition we introduce the new quantity
\begin{equation}
\label{gamma} \gamma:=\frac{(\Delta p)^2}{4p_0(1-p_0)\tau}\,,
\end{equation}
which will turn out to be the {\sl decoherence rate}.
It contains apart from $\Delta p$ and $p_0$ also the time
interval $\tau$ between two measurements. These three parameters
characterize completely the influence of the sequence of pure
measurements. $\gamma$ increases when the measurements become
sharper and accordingly have a stronger influence on the qubit. A
decreasing time difference $\tau$ between two measurements
results as well in an increase of $\gamma$. This reflects a Zeno
type effect which also happens for unsharp measurements.

Installing $\gamma$ we get the ultimate form of the operations,
valid for $\Delta p\ll\tau/ \Delta t\ll1$:
\begin{equation}
\label{Ms} |M_s|= \frac{1}{\sqrt[4]{2\pi/(\gamma\Delta t)}}
\exp\bigg\{-\gamma\frac{(\sigma_z -s)^2}{4} \Delta t\bigg\}\,.
\label{diffeq}
\end{equation}
The resulting effects $E_s = |M_s|^2$ have Gaussian form. They
show the characteristics which are known for instance from an
unsharp position measurement as investigated e.g. in
\cite{CavesMilburn87}. Instead of a continuous observable
however, we are dealing here with a discrete observable.

With reference to $s$, the complete operation element including
the back-action and the \lq\lq free"  evolution is given by
\begin{equation}
\label{UMs} M_s=U_s|M_s| \exp \Bigl\{ - \frac{i}{\hbar} H \Delta
t \Bigr\} \,,
\end{equation}
where, using (\ref{UN}) and (\ref{defs}), we obtain for $U_s$
which replaces $U(N_+, N)$:
\begin{equation}
\label{Us} U_s=\exp{\left\{-\frac{i}{\hbar}H_{AV}\Delta t
-\frac{i}{2\hbar}\Delta H s\Delta p\Delta t\right\}}\,.
\end{equation}
We have thereby introduced the {\sl averaged back-action Hamiltonian}
$H_{AV}$ and the difference $\Delta H$ of the back-action
Hamiltonians, respectively:
\begin{eqnarray}
H_{AV}&:=&p_0H_++(1-p_0)H_-\\
\Delta H&:=&(H_--H_+)\,.
\end{eqnarray}
$M_s$ of (\ref{UMs}) replaces $M(N_+,N)$ of eqn.\ (\ref{M}) for the
continuous variable $s$.

The N-series operation elements above correspond to a continuous
set of effects with the Gaussian distribution function
\begin{equation}
\label{ps} p_s=\langle M_s^+ M_s\rangle_\rho\ .
\end{equation}
The completeness relation is satisfied if we extend the range of
$s$ to the whole real axis:
\begin{equation}
\label{norm} \int_{-\infty}^{\infty}\:M_s^\dagger M_s\
ds=1\hspace{-1.57mm}1\,.
\end{equation}
The statistical weight of the unphysical values of $s$ will be
negligible provided that $p_0\not\approx 0\,,\,1$ and this  justifies the formal extension of the values
of $s$ beyond their physical range (\ref{lims}).

\section{Non-selective evolution}
In the non-selective case the state change during a N-series can
be expressed in the operator-sum representation  as
\begin{equation}
\rho(t+\Delta t) = \int_{-\infty}^{\infty}\,M_s \rho
 M_s^+ ds\,.
\end{equation}
We are going to expand the r.h.s. up to  linear terms in $\Delta
t$.

The unitary parts of the operation $M_s$ which are generated by
$H$ , $H_{AV}$ and $\Delta H$ lead to
\begin{equation}\label{Uexpand}
\Delta \rho:= \rho(t+\Delta t)- \rho(t) = -\frac{i}{\hbar}[\,H +
H_{AV}\,,\,\rho (t)\,]\Delta t + D(\rho (t)) - \rho (t)\,.
\end{equation}
The integral
\begin{equation}
D(\rho):= \int_{-\infty}^\infty
\exp{\left\{-\frac{i}{2\hbar}\Delta H s\Delta p\Delta
t\right\}}|M_s|\rho|M_s|\,\exp{\left\{\frac{i}{2\hbar}\Delta H
s\Delta p\Delta t\right\}}\,ds\,
\end{equation}
over the parts which depend on $s$ will first be calculated and
then expanded.

Introducing operators which act from the left and are denoted with
L (e.g. $\sigma_z^{\mbox{L}}\rho:= \sigma_z\,\rho$) as well as
operators which act from the right and are denoted with R (e.g.
$\sigma_z^{\mbox{R}}\rho:= \rho\,\sigma_z$), we rewrite the
integrand of $D(\rho)$
\begin{eqnarray}
D (\rho)& = &\frac{1}{\sqrt{2\pi\gamma/\Delta
t}}\int_{-\infty}^\infty\,\exp{\left\{-\frac{i}{2\hbar}
\left(\Delta H^{\mbox{L}}-\Delta H^{\mbox{R}}\right) s\Delta p\Delta
t\right\}}\nonumber\\
&&\times
\exp\left\{-\frac{\gamma}{4}\left((\sigma_z^{\mbox{L}}-s)^2
+(\sigma_z^{\mbox{R}}-s)^2\right) \Delta t\right\} \,ds\,\rho \,.
\label{drhoexpand}
\end{eqnarray}
It is important to take into account a further operator ordering
for the integrand, namely that $\Delta H^{\mbox{L}},\Delta
H^{\mbox{R}}$ should remain leftmost and rightmost, respectively.
The resulting integral is Gaussian in $s$. It may be solved in a
closed form:
\begin{eqnarray}
D(\rho)& =& \exp\biggl\{-\biggl[\frac{\gamma}{8}(\sigma_z^{\mbox{L}}-
\sigma_z^{\mbox{R}})^2
+ \frac{\Delta p^2}{8\gamma\hbar^2}\left(\Delta
H^{\mbox{L}}-\Delta H^{\mbox{R}}\right)^2 \nonumber\\
 &&+i\frac{\Delta p}{4\hbar}\left(\sigma_z^{\mbox{L}}+\sigma_z^{\mbox{R}}\right)
\left(\Delta H^{\mbox{L}}-\Delta H^{\mbox{R}}\right) \biggr]
\Delta t\biggr\} \, \rho\,.
\end{eqnarray}

Now we  expand it up to the leading linear term in $\Delta t$ and
 restore the usual operator formalism according to the rules for
 $L$ and $R$. This leads for example to
\begin{eqnarray}
(\sigma_z^L - \sigma_z^R )^2 \rho
&=&(\sigma_z^L-\sigma_z^R)(\sigma_z^L-\sigma_z^R) \rho \nonumber\\
&=&(\sigma_z^L-\sigma_z^R)[\sigma_z,\rho]=\bigl[\sigma_z,[\sigma_z,\rho]\bigr]
\end{eqnarray}
and all together to
\begin{eqnarray}
\label{Drho}
&&D(\rho)-\rho = \\
&&\left(- \frac{\gamma}{8}
[\,\sigma_z\,,\,[\,\sigma_z\,,\,\rho\,]\,] -\frac{(\Delta
p)^2}{8\gamma\hbar^2}[\Delta H,[\Delta H,\rho]\,] -i\frac{\Delta
p}{4\hbar}[\Delta H,\{\sigma_z,\rho\}\,]\right)\Delta t\,.
\nonumber
\end{eqnarray}
%where we have omitted terms of order $O(\Delta t^2\gamma^2)$ and of order
%$\mathcal{O}(\Delta p\Delta H \Delta t/\hbar)$,  as well as terms of
%order $\mathcal{O}(\Delta p^2\Delta H^2 \Delta t/\hbar^2)$ or smaller.
While the second and the third term are proportional to small
quantities (cp.\ (\ref{eqN}) -- (\ref{Hfeedsmall})), the first
contains the ratio of the two small quantities $\Delta p^2$ and
$\tau$ (cp. (\ref{gamma})). We assume $\gamma$ not to be small.
We thus obtain as final result for the state evolution during one
N-series
\begin{equation}\label{diffmaster}
\Delta \rho
=\Bigl(\,-\frac{i}{\hbar}\,[\,H+H_{AV}\,,\,\rho\,]
-\frac{\gamma}{8}\,[\,\sigma_z\,,\,[\,\sigma_z\,,\,\rho\,]\,]
\,\Bigr)\Delta t\,.
\end{equation}
The first term on the r.h.s.\ represents the unitary dynamical
evolution related to the \lq\lq free"  Hamiltonian $H$ and to the
averaged Hamiltonian $H_{AV} = p_0 H_++(1-p_0)H_-$. $p_0$
and $1-p_0$ are approximately the probabilities that the
back-action causes a Hamiltonian development with $H_+$ or $H_-$
respectively (cp.\ section \ref{sectwo}). The second term on the
r.h.s. reflects the decoherence induced by the pure measurement part
$|M_\pm|$ of  the operation. The structure of both terms is clearly
what one would expect on physical grounds. 

The second and third term of eqn.\ (\ref{Drho}) indicate additional
physical effects, which are to be expected in a higher order
approximation. The second term 
corresponds to further decoherence induced by unitary
back-action. The third term goes back to friction caused also by
the back-action. A complete list of additional terms is given in
Appendix C

Eqn.\ (\ref{diffmaster}) has been derived on the basis of the
following approximations: We have changed the order of $U$,
$U_\pm$ and $|M_\pm|$ in the operation of the N-series,
neglecting the commutators between them, cp.\ (\ref{C1}) and
(\ref{C2}), the estimated error is smaller than
$\mathcal{O}(N\Delta p \Delta t \Vert H \Vert /\hbar) +
\mathcal{O}(\Delta t^2 \Vert H \Vert \max\{\Vert
H_\pm\Vert\}/\hbar^2)$ $+\mathcal{O}(N\Delta p \Delta t \Vert
\max\{\Vert H_\pm\Vert\}/\hbar)$ $+ \mathcal{O}(\Delta t^2 \Vert
H_+\Vert\Vert H_-\Vert/\hbar^2)$. We also have approximated the
q-number denominator of the Gaussian operation elements by a
c-number, cp. (\ref{EplusEmin}), which leads to an error of order
$\mathcal{O}(\Delta p \Delta t\gamma)$. We further expanded the
operation in powers of $\Delta t $  up to the first order, cf.
(\ref{Uexpand}) and (\ref{drhoexpand}), which results in errors
of order $ O(\Delta t^2 \Vert H+H_{AV} \Vert^2/\hbar^2)$ and
$O(\Delta t^2 \gamma \Vert H+H_{AV} \Vert/\hbar)$.
%Finally we omitted terms of order
%$\mathcal{O}(\Delta p\Delta H \Delta t/\hbar)$ and $\mathcal{O}(\Delta
%p^2\Delta H^2 \Delta t/\hbar^2)$ or smaller.
In the continuum limit the errors all vanish but they can play an
important role for discrete sequences of measurements if $\Delta
p \ll 1$ is not fulfilled. Appendix C contains a calculation of
the state change up to higher orders.
% where some additional terms with a well
%known physical interpretation such as decoherence and dissipation appear.
This more accurate calculation confirms the order of the errors
estimated here. In case of the neglected commutators in
(\ref{C1}) and (\ref{C2}) the error turns out to be actually
smaller (cp.\ (\ref{exact})).

\section{Continuum limit}
In Section 4 we have worked out the operation given by the
discrete state transformation between an initial state $\rho(t)$
and the final state $\rho(t+ \Delta t)$  after a N-series of
instantaneous interactions of a qubit with an environment, which
are of the type of an unsharp measurement. It can be applied to a
truly sequential measurement by dividing the sequence of
elementary measurements into a succession of N-series. Since the
r.h.s. of the equation (\ref{diffmaster}) is proportional to
$\Delta t$, the given approximation is not sensitive to the
division as long as $N$ is large. The rate of the change of
$\rho$ is invariant under this the division. This discrete-time
analysis is the most natural approach to sequential measurements.
For example see \cite{AudretschKonradScherer01}. Eqn.\
(\ref{diffmaster}) above reveals the underlying physics as
represented by the decoherence rate $\gamma$ and the measurement
induced unitary development given by $H_{AV}$.

There are elaborated schemes for the treatment of permanently open
quantum systems by continuous-time descriptions. Master equations
are an example. One may profit from these schemes as
approximations in the sequentially open case too, if 
a physically reasonable continuum limit
$\tau\rightarrow 0$ is carried out. 
The corresponding demand for such a limit is
that the physical characteristics of the sequential situation
have to be taken over. We proceed as follows:

The quantity $p_0$ is the mean probability to obtain the
measurement result $+$. We leave the value of $p_0$ unchanged in
the continuum limit. In order to not change the decoherence
behavior in the continuum limit we secondly demand for the
decoherence rate:
\begin{equation}
\label{limgamma}
\lim_{\tau\rightarrow 0} \frac{(\Delta p)^2}{4p_0(1-p_0)\tau} = \gamma= \mbox{const}\,.
\end{equation}
The smaller $\Delta p$ the weaker the single measurement. With
$\tau \rightarrow 0 $ and the strength $\Delta p$ of the single
measurement unchanged, a Zeno effect would be obtained. This is
prevented by appropriately diminishing the strength $\Delta p$ of
the measurement according to (\ref{limgamma}). This demand can
also be found in the literature \cite{Barchielli82}.

If in a given sequential physical situation the $H_{AV}$
 is non-vanishing, then the total Hamiltonian dynamics
is according to (\ref{diffmaster}) governed by the Hamiltonian
$H+H_{AV}$. We want to keep this dynamics in the continuum limit
on physical grounds and demand therefore that $H_{AV}$ remains
unchanged. Performing the limit $\tau\rightarrow 0$ as specified
above results in the master equation
\begin{equation}
\label{master} \dot{\rho}=-\frac{i}{\hbar}\, [\,H+
H_{AV}\,,\,\rho\,]-\frac{\gamma}{8}\,
[\,\sigma_z\,,\,[\,\sigma_z\,,\,\rho\,]\,]\,,
\end{equation}
%(\ref{master}) is still only an approximation because we assumed $N$
%to be large but fixed. In order to make (\ref{MN}) and (\ref{Ms0}) the
%exact operation elements of an N-series, we have to assume for $\tau\rightarrow 0$
%\begin{equation}
%N\rightarrow \infty\,,\, \tau \rightarrow 0\quad \mbox{with}\,\,
%\Delta t= N\tau\rightarrow 0\,\,\mbox{and}\,\, N\Delta p\rightarrow 0\,.
%\end{equation}
%(cp. (\ref{condH},\ref{Hfeedsmall}, \ref{Ndeltap}) and
%(\ref{condN}).
which describes approximately the discontinuous situation in the
noisy channel characterized above.

We note that the master equation (\ref{master}) could have been
obtained in the limit (\ref{limgamma}) directly from the
elementary measurements. The intervening formulation in terms of
N-series will be exploited by the forthcoming equations of
selective evolution.

\section{Selective evolution}
In Sec. 3 we calculated the Gaussian form (\ref{Ms}) of effective
operation (\ref{UMs}) valid for a N-series. In Sec. 5 we derived
the master equation (\ref{master}) valid exactly in the continuous
limit (\ref{limgamma}). As a matter of fact, the master equation
describes the non-selective evolution. Selective evolution is, on
the contrary, conditioned on the random measurement results
(readout) and described by stochastic equations. In our case, the
readout is $s$. It is the continuously measured unsharp value of
the observable $\sigma_z$ obtained in the N-series in the limit
(\ref{limgamma}).

The theory of the selective evolution has been available since
long ago \cite{Diosi88}. From the Gaussian operations (\ref{Ms})
in the limit $\Delta t\rightarrow0$, it has been proved that
the selective evolution of the quantum state, conditioned on the
measurement result $s$, satisfies the conditional master equation:
\begin{equation}\label{Itorho}
\dot\rho=-\frac{i}{\hbar}[H+H_{AV},\rho]
         -\frac{\gamma}{8}[\sigma_z,[\sigma_z,\rho]\,]
      +w\frac{\sqrt{\gamma}}{2}\{\sigma_z-\langle\sigma_z\rangle,\rho\}
\end{equation}
The function $w(t)$ is the standard white-noise and the equation
should be understood in the Ito-stochastic sense.
The state evolution
couples to the readout $s$ by:
\begin{equation}\label{Itos}
s=\langle\sigma_z\rangle+\frac{1}{\sqrt{\gamma}}w
\end{equation}
Obviously the stochastic mean of the conditional master equation
(\ref{Itorho}) reduces to the unconditional master equation
(\ref{master}) as it should. Of course eqn.\ (\ref{Itorho}) applies
to pure initial states as well. Then the pure state property
$\rho^2=\rho$ is preserved. The derivation may completely be
identical to that in Ref.~\cite{Diosi88}. In the continuum limit
(\ref{limgamma}) the value of $\Delta p$ must vanish and the
feed-back $U_s$ is thus deterministic, given by $H_{AV}$ alone.

The above equations of selective evolution are exact in the
following sense. Elementary operations are being applied with
frequency growing to infinity and strength decreasing to zero as
given by (\ref{limgamma}), i.e. at fixed $\gamma$. We read out
the rate $N_+/N$ averaged over time $\Delta t$ which should go to
zero in such a way that $N=\Delta t/\tau$ still goes to infinity.
The elementary time $\tau$ goes \lq\lq faster"  to zero than the time
$\Delta t$ to calculate the rate $N_+/N$. The calculated current
rate $N_+/N$ is related to $s$ by (\ref{defs}):
\begin{equation}
s=\frac{p_0-N_+/N}{\sqrt{\gamma p_0(1-p_0)\tau}}
\end{equation}
The continuous limit (\ref{limgamma}) of $s$ exists. As follows from
(\ref{Itos}), it is centered around a state dependent part
$\langle\sigma_z\rangle$ and superposed by
the white-noise of constant intensity $1/\gamma$.

\goodbreak

\bigskip
\noindent {\bf Appendix A}
\medskip

\noindent In the appendices we sketch the calculation of the
change of state in the non-selective regime including all terms
up to order $\mathcal{O}(\Delta t \Delta p^2)$ and
$\mathcal{O}(\Delta t^2)$, where $\Delta t$ occurs in products
with either $H/\hbar, H_\pm/\hbar$ or $\gamma$.

We start with the exact operation element $\Omega$ for a N-series
with unitary development $U$ between consecutive measurements.
\begin{equation}
\Omega_{(m_i)}:=
U_{m_N}\,...\,U_{m_1}|M_{m_N}|\,...\,|M_{m_1}|U^N + R_1 + R_2\,,
\end{equation}
where $R_1$ and $R_2$ are the terms which arise from commuting out the
evolution operators $U$ and the feedback operators $U_\pm$
respectively, cp. equations (\ref{C1}) and (\ref{C2}). The
relation of $R_1$ and $R_2$ to $C_1$ and $C_2$ and an estimation of the
order of magnitude of  $C_1$ and $C_2$ is described below.
Since the commutators $K_{m_i}:= [ \,U\,,\,M_{m_i}]$ occurring in $R_1$
are of order
$\mathcal{O}(\Delta p\tau \Vert H\Vert/\hbar) +
\mathcal{O}(\tau^2\Vert H\Vert
\max\{\Vert H_\pm\Vert \}/\hbar^2)$, we can neglect terms containing
products of two such commutators.
\begin{eqnarray}
\label{R1}
R_1 &=& M_{m_N}K_{m_{N-1}}\,...\,M_{m_1} U^{N-1} + 2
M_{m_N}M_{m_{N-1}}K_{m_{N-2}}\,...\,M_{m_1} U^{N-1}\nonumber\\
&&+\,...\,+ (N-1) M_{m_N}\,...\,K_{m_1} U^{N-1}\\
&=& \Delta p
\tau[\,H\,,\,\sigma_z\,]\,U_{m_N}\,...\,U_{m_1}\sum_{n=1}^{N-1}\,\frac{n}{a_{m_{N-n}}}
|M_{m_N}|\,...\,|M_{m_1}|U^{N-1}\nonumber\\
& -& \frac{\tau^2}{\hbar^2}\ \sum_{n=1}^{N-1}\,n
\,[\,H\,,\,H_{m_{N-n}}\,]\,U_{m_N}\,...\,U_{m_1}\,
|M_{m_N}|...|M_{m_{N-n}}|...|M_{m_1}|U^{N-1}\,,\nonumber
\end{eqnarray}
where $a_+:=-4\sqrt{p_0}$ and $a_-:=4\sqrt{1-p_0}$. Please note,
that in (\ref{R1}) behind the
sum sign the products of  $|M|$'s and $U$'s are meant to not contain
$M_{m_{N-n}}$ and $U_{m_{N-n}}$,
except if explicitly mentioned.
In the last two lines we have commuted out the feedback $U_\pm$. The
resulting error
is of higher order and can be neglected.

Also in $R_2$ we only take into account the terms containing one
commutator $K_{k,l}:= [\,M_{m_k}\,,\,U_{m_l}\,]$, which is of order
$\mathcal{O}(\Delta p\tau\max\{\Vert H_\pm\Vert \}/\hbar)$.

\begin{eqnarray}\label{R2}
R_2 &=& U_{m_N}\,...\,U_{m_2}K_{2,1}|M_{m_N}|\,...\,|M_{m_3}||M_{m_1}| U^{N}+\,...\,\\
&&+ \Biggl(U_{m_N}\,...\,U_{m_2}K_{N,1}+
U_{m_N}\,...\,U_{m_3}U_{m_1}K_{N,2}+\,...\nonumber\\
&&+ \,U_{m_N}U_{m_{N-2}}\,...\,U_{m_1}K_{N,N-1}\Biggr)
|M_{m_{N-1}}|\,...\,|M_{m_1}| U^{N}\nonumber\\
&=& -\frac{i\Delta p
\tau}{\hbar}\sum_{k=2}^{N}\sum_{l=1}^{k-1}\,\frac{1}{a_{m_{k}}}[\,\sigma_z\,,\,H_{m_l}\,]\,U_{m_N}\,...\,U_{m_{l+1}}U_{m_{l-1}}\,...\,
U_{m_1}\times\nonumber\\
&&\times \,|M_{m_N}|\,...\,|M_{m_{k+1}}||M_{m_{k-1}}|\,...\,|M_{m_1}|U^{N-1}\nonumber\\
&& -\frac{\tau^2}{\hbar^2}\sum_{k=2}^{N}\sum_{l=1}^{k-1}\,
[\,H_{m_k}\,,\,H_{m_l}\,]\,U_{m_N}\,...\,U_{m_{k+1}}U_{m_{k-1}}\,\,...\,U_{m_{l+1}}\times\nonumber
\\
&& \times \,U_{m_{l-1}}\,...\,
U_{m_1}|M_{m_N}|\,...\,|M_{m_1}|U^{N-1}\nonumber\,.
\end{eqnarray}
Let us shortly motivate the estimation of the order of magnitude of
$C_1$ and $C_2$ given in (\ref{NormC1}) and (\ref{NormC2}). First
we observe that $R_1= M_{m_n}\,...\,M_{m_1}U^N C_1$ and
$R_2= U_+^{N_+}U_-^{N-N_+}|M_+|^{N_+}|M_-|^{N-N_+}U^N C_2$. A moment's
thought shows that the order of magnitude of the summands contained in
$C_i$ is equal to the order of the commutators $K_{m_i}$ and $K_{k,l}$
in $R_i$. Since there are
approximately $N^2$ such summands in $C_i$, the norm of $C_i$ can be estimated
to be less or equal to $N^2$ times the order of the  commutators in
$R_i$ which leads to the claims (\ref{NormC1}) and (\ref{NormC2}).

\bigskip
\noindent{\bf Appendix B}
\medskip

\noindent The state change due to a N-series in the non-selective
regime reads
\begin{eqnarray}
\label{nonselect}
\rho(t+\Delta t)& = & \sum \Omega_{(m_i)}\,\rho\,\Omega_{(m_i)}^+\nonumber\\
&=&\sum_{m_1\,...\,m_N}
U_{m_N}\,...\,U_{m_1}|M_{m_N}|\,...\,|M_{m_1}|U(\Delta t)\,\times
\nonumber\\
&&\times\rho\,
U^+(\Delta t)|M_{m_1}|\,...\,|M_{m_N}|U_{m_1}^+\,...\,U_{m_N}^+\nonumber\\
&&+ \tilde{R_1}+ \tilde{R_2} + \mathcal{O}(R_1^2)+
\mathcal{O}(R_2^2) + \mathcal{O}(R_1R_2)
\end{eqnarray}
with
\begin{equation}
\label{R12}
\tilde{R_i}:= \sum_{m_1\,...\,m_N}\left\{R_i\,\rho\,U^+(\Delta
t)|M_{m_1}|\,...\,|M_{m_N}|U_{m_1}^+\,...\,U_{m_N}^+ + h.c.\right\}\quad i=1,2\,.
\end{equation}
In order to carry out the summation in (\ref{nonselect}) its terms can
be expressed by means of binomial distributions. The latter can be
approximated by integrals over Gaussians. For the first term in
(\ref{nonselect}) this recipe has already been demonstrated in section
3 and 4. $\tilde{R_1}$  can be written in binomial
form by observing that:
\begin{eqnarray}
\label{f1}
&&\sum_{m_1\,...\,m_N}\sum_{n=1}^{N-1}\,n\,
b_{m_{N-n}}U_{m_N}\,...\,U_{m_1} |M_{m_N}|\,...\,|M_{m_{N-n+1}}|
|M_{m_{N-n-1}}|\,...\,|M_{m_1}|\times \nonumber\\
&& \quad\times\rho\ |M_{m_1}|\,...\,|M_{m_N}|
U_{m_1}^+\,...\,U_{m_N}^+ \nonumber\\
&&=
\frac{(N-1)(N-2)}{2}\left(\sum_m\,b_mU^L_m|M_m|^R (U_m^+)^R
\right)
\sum_{N_+=0}^{N-1}{N-1\choose N_+}U_+^{N_+}
U_-^{N-N_+-1}\,\times\nonumber \\
&&\times|M_+|^{N_+}|M_-|_-^{N-N_+-1}\,\rho\,
|M_+|^{N_+}|M_-|_-^{N-N_+-1}(U_+^+)^{N_+}(U_-^+)^{N-N_+-1}\,,
\end{eqnarray}
where we have again used the notation that operators with upper script
$L$ and $R$ act from the left
and from the right respectively. A similar formula is obtained when instead of
$|M_{m_{N-n}}|$ in the first line $U_{m_{N-n+1}}$ is missing. Then
only $U_m^L$  has to be replaced by  $|M_m|^L$.

$\tilde{R}_2$ can be
simplified employing
\begin{eqnarray}
\label{f2}
&&\sum_{m_1\,...\,m_N}\sum_{k=2}^{N}\sum_{l=1}^{k-1}\,C_{m_k,m_l}U_{m_N}\,...\,U_{m_{k+1}}
U_{m_{k-1}}\,...\,U_{m_1}\times\\
&&\times |M_{m_N}|\,...\,|M_{m_{l+1}}|
|M_{m_{l-1}}|\,...\,|M_{m_1}|\,\rho\ |M_{m_1}|\,...\,|M_{m_N}|
U_{m_1}^+\,...\,U_{m_N}^+ \nonumber\\
&& =
\frac{(N-1)(N-2)}{2}\left(\sum_{m,\acute m=+,-}C_{m,\acute
m}U_m^L|M_{\acute m}|^L(U_m^+)^R |M_m|^R(U_{\acute m}^+)^R |M_{\acute m}|^R
\right)\times\nonumber\\
&&\times\sum_{N_+=0}^{N-2}{N-2\choose N_+}U_+^{N_+}
U_-^{N-N_+-2}|M_+|^{N_+}
|M_-|^{N-N_+-2}\,\times \nonumber\\
&&\times\rho\,|M_+|^{N_+}|M_-|^{N-N_+-2}(U_+^+)^{N_+}(U_-^+)^{N-N_+-2}\,,\nonumber
\end{eqnarray}
 Formulae (\ref{f1}) and
(\ref{f2}) neglect commutators between the operators they contain. In
our case corrections containing these commutators would be of higher
order and therefore too small.

Applying formulae (\ref{f1}),
(\ref{f2}) to $\tilde{R_1}$, $\tilde{R_2}$  respectively and expressing $N_+/N$ in terms of variable $s$
according to (\ref{defs}) we obtain:
\begin{eqnarray}
\tilde{R_1}&=& \frac{-iN\Delta p\Delta
t}{2\hbar}\left\{[\,H\,,\,\sigma_z\,]\, \sum_m\,
\frac{U_m}{a_m}\tilde{D}(\rho) |M_m|U^+_m +h.c.\right\}\\
&&-\frac{\Delta t^2}{2\hbar^2}\sum_m\left\{
\,[\,H\,,\,H_m\,]\,|M_m|\,\tilde{D}(\rho)\,|M_{m}|U_m^+
+h.c.\right\}\,,\nonumber
\end{eqnarray}
where $\tilde{D}(\rho)= \int_{-\infty}^{\infty}M_s\,\rho\,M_s^+ ds$
with $M_s$ as given by (\ref{UMs}) with $\Delta t$ replaced by $(N-1)\tau$.
$\tilde{R_2}$ now reeds
\begin{eqnarray}
\label{tildeR2}
\tilde{R_2}&=& -\frac{iN\Delta p\Delta t}{2\hbar}\sum_{m,\acute
m}\,\left\{\frac{[\,\sigma_z\,,\,H_{\acute m}\,]}{a_m}U_m|M_{\acute
m}|\tilde{D}(\rho)|M_{\acute m}|U_{\acute m}^+|M_m|U_m^+
+h.c.\right\}\nonumber\\
& -&\hspace{-0.3cm}\frac{\Delta t^2}{2\hbar^2}\sum_{m\not=\acute m}\,\left\{
[\,H_{m}\,,\,H_{\acute m}\,]\,|M_m||M_{\acute
m}|\tilde{D}(\rho)|M_{\acute m}|U_{\acute m}^+|M_m|U_m^+
+h.c.\right\}
\end{eqnarray}
In $\tilde{R_2}$, when inserting $M_s$ from (\ref{UMs}) in
$\tilde{D}(\rho)$,
$\Delta t$ has to be replaced by $(N-2)\tau$.
The second sum in (\ref{tildeR2}) vanishes since the summand with $m=+\,,\,\acute m=-$ and
the summand with $m=-\,,\,\acute m=+$ add to zero.
Inserting the lowest order of $\tilde{D}(\rho)$ namely
$\tilde{D}(\rho)\approx \rho$, it is easy to see that $\tilde{R_1}$
and $\tilde{R_2}$ contribute the terms in the sixth   and the
seventh line of equation (\ref{exact}) respectively to the change of
state.

\bigskip
\noindent{\bf Appendix C}
\medskip

\noindent Having calculated $\tilde{R_1}$ and $\tilde{R_2}$, we
want to sketch how to process the main contribution to the state
change, which is represented by the first term in equation
(\ref{nonselect}). As mentioned above this part of the operation
can be written by means of a binomial distribution and then be
expressed with operation elements whose modulus $|M(N_+,N)|$ are
the square root of Gaussians, cf. equation (\ref{MN}). In
contrast to section 3 and 4, we now take into account the full
q-number denominators of the modulus $|M(N_+,N)|$ in (\ref{MN}).
Expressing the operation elements in terms of variable s (cp.
(\ref{defs})) we obtain for their modulus
\begin{equation}
\label{MsA}
|M_s|=
\frac{1}{\sqrt[4]{2\pi/(\hat \gamma\Delta t)}}
\exp\bigg\{-\hat\gamma\frac{(\sigma_z -s)^2}{4}\Delta t\bigg\}\,,
\label{diffeqA}
\end{equation}
with
\begin{equation}
\label{gammaA}
\hat \gamma:=\frac{(\Delta p)^2}{4E_+E_-\tau}
\end{equation}
Expanding the unitary part of the operation up to order $\Delta t^2$
leads to the state change (without the contribution from $R_1$ and $R_2$)
\begin{eqnarray}
\label{expU}
\rho( t+\Delta t) &=& D(\rho) -\frac{i\Delta t}{\hbar}\,[\,H+H_{AV}\,,\,D(\rho)\,]\:-\frac{\Delta
t^2}{2\hbar^2}\,\{\,(H+H_{AV})^2\,,\,\rho\,\}\nonumber\\
&&+\frac{\Delta t^2}{\hbar^2}(H+H_{AV})\,\rho\,(H+H_{AV})
\end{eqnarray}
with
\begin{eqnarray}
\label{Drhoint}
D(\rho)&=& \int_{-\infty}^\infty\exp
\left\{\frac{i}{2\hbar}\left(\Delta H^{\mbox{L}}-\Delta H^{\mbox{R}}\right)s\Delta p \Delta
t\right\}
\exp\left\{-\frac{\hat \gamma^{\mbox{L}}}{4}\left(\sigma_z^{\mbox{L}}-s\right)^2\Delta
t\right\}\times \nonumber\\ &&
\times\,\exp\left\{-\frac{\hat \gamma^{\mbox{R}}}{4}\left(\sigma_z^{\mbox{R}}-s\right)^2\Delta
t\right\}ds
\frac{\sqrt[4]{\hat \gamma^{\mbox{L}}\hat \gamma^{\mbox{R}}}}{\sqrt{2\pi/\Delta t}} \,\rho\,.
\end{eqnarray}
We note that in eqn.\ (\ref{expU}) $H$ is meant to act in operator
products directly on $rho$. This is due to the order of
operators in the operation elements (cp.\ (\ref{Us})).
The integral $D(\rho)$ has a closed form solution which can be expanded in powers
of $\Delta t$ and $\Delta p$. $\gamma$ without hat is given by
(\ref{gamma}).
\begin{eqnarray}
D(\rho)& =& \Biggl\{1 -\Delta t\biggl[\frac{\gamma}{8}(\sigma_z^{\mbox{L}}-
\sigma_z^{\mbox{R}})^2\biggl(1-
\frac{1}{2}\left(\frac{\Delta p
(p_0-1/2)}{p_0\tilde{p_0}}\right)^2\times\\
&& \times \left(1-\sqrt{p_0\tilde{p_0}}(p_0 \tilde{p_0}+ 3-2^{-1/4})\right)\biggr)\nonumber\\
&&+ \frac{\Delta p^2}{8\gamma\hbar^2}\left(\Delta
H^{\mbox{L}}-\Delta H^{\mbox{R}}\right)^2
+i\frac{\Delta p}{4\hbar}(\Delta H^{\mbox{L}}-\Delta H^{\mbox{R}})(\sigma_z^{\mbox{L}}+\sigma_z^{\mbox{R}})\,
\biggr]\nonumber\\
&& + \Delta t^2\frac{\gamma^2}{32}\,(\sigma_z^{\mbox{L}}-
\sigma_z^{\mbox{R}})^2
+ \mathcal{O}(\Delta p \Delta t^2)
+ \mathcal{O}(\Delta p^2 \Delta t^2) + \mathcal{O}(\Delta t^3)\Biggr\}
\, \rho\,.\nonumber
\end{eqnarray}

Collecting all terms up to  order
$\mathcal{O}(\Delta t^2)$ and $\mathcal{O}(\Delta t \Delta p^2)$ we
obtain the following difference equation:
\begin{eqnarray}
\label{exact}
\Delta \rho
&=&\Delta
t\Biggl(-\frac{i}{\hbar}\,[\,H+H_{AV}\,,\,\rho\,]\\
&&-\:\frac{\gamma}{8}\left(1-
\frac{1}{2}\left(\frac{\Delta p
(p_0-1/2)}{p_0\tilde{p_0}}\right)^2\left(1-\sqrt{p_0\tilde{p_0}}(p_0 \tilde{p_0}+ 3-2^{-1/4})\right)\right)
\,[\,\sigma_z\,,\,[\,\sigma_z\,,\,\rho\,]\,]\nonumber\\
&&-\frac{\Delta p^2}{8\gamma \hbar^2}\,[\,\Delta H\,,\,[\Delta
H\,,\,\rho\,]\,]\,
- \frac{i\Delta
p}{4\hbar}[\Delta H,\{\sigma_z,\rho\}\,]\Biggr)\nonumber\\
&&+\Delta t^2
\Biggl(-\frac{i\gamma}{8\hbar}\bigl([\,\sigma_z\,,\,[\,\sigma_z\,,\,[\,H\,,\rho]\,]\,]+
[\,H_{AV}\,,\,[\,\sigma_z\,,\,[\,\sigma_z\,,\,\rho\,]\,]\,]\bigr)\nonumber\\
&&
-\frac{1}{2\hbar^2}\{\,(H+H_{AV})^2\,,\,\rho\,\}\, +\frac{1}{\hbar^2}(H+H_{AV})\rho(H+H_{AV})
+\frac{\gamma^2}{32}\,[\,\sigma_z\,,\,[\,\sigma_z\,,\,\rho\,]\,]\Biggr)
\nonumber\\
&&\left(-\frac{i\Delta t^2\gamma}{2\hbar}+\frac{3i\Delta t \Delta p^2}
{8\hbar p_0\tilde{p_0}}\right)\Biggl(\biggl\{[\,H\,,\,\sigma_z\,]\,\rho\,\sigma_z
+ \mbox{h.c.}\biggr\}
-\frac{i}
{\hbar}[\,[\,H,\,\,H_{AV}\,]\,,\,\rho\,]\nonumber\\
&&
+\frac{1}{4}\biggl\{[\,\sigma_z\,,\,\tilde{p_0} H_+
+p_0 H_-\,]\,\rho\,\sigma_z\,+ \mbox{h.c.}\biggr\}\Biggr)\nonumber\\
&& \mathcal{O}(N\Delta p \Delta t^2) + \mathcal{O}(\Delta p \Delta t^2)
+ \mathcal{O}(\Delta p^2 \Delta t^2) + \mathcal{O}(\Delta t^3)\nonumber
\end{eqnarray}
with $\tilde{p_0}:=1-p_0$. In the order terms $\Delta t$ occurs in
products with one of the three:  $H/\hbar, H_\pm/\hbar$ or $\gamma$.
%\bibliography{measurement}
%\bibliographystyle{unsrt}

\end{document}